\documentclass[aps,pre,twocolumn,showpacs,showkeys]{revtex4}
\usepackage{umlaut}
\usepackage{graphicx}

\begin{document}

\title{Influence of external flows on crystal growth: numerical investigation}

\author{Dmitry Medvedev${}^{1,2}$, Thomas Fischaleck${}^{1}$,
  Klaus Kassner${}^{1}$ \\
  ${}^{1}$ Institut für Theoretische Physik, \\
  Otto-von-Guericke-Universität
  Magdeburg, Germany\\
  ${}^{2}$ Institute of Hydrodynamics of the Russian Academy of
  Sciences, \\ Novosibirsk, Russia}

%\centerline{\bf Abstract}
\begin{abstract}

We use a combined phase-field/lattice-Boltzmann scheme
%\cite{dmitryjcg05,MKas2005b}
[D. Medvedev, K. Kassner, Phys. Rev. E {\bf 72}, 056703 (2005)] to
simulate non-facetted crystal growth from an undercooled melt in
external flows. Selected growth parameters are determined numerically.

For growth patterns at moderate to high undercooling and relatively
large anisotropy, the values of the tip radius and selection parameter
plotted as a function of the P\'{e}clet number fall approximately on
single curves. Hence, it may be argued that a parallel flow changes
the selected tip radius and growth velocity solely by modifying
(increasing) the P\'{e}clet number. This has interesting implications
for the availability of current selection theories as predictors of
growth characteristics under flow.

At smaller anisotropy, a modification of the morphology diagram in the
plane undercooling versus anisotropy is observed.  The transition line
from dendrites to doublons is shifted in favour of dendritic patterns,
which become faster than doublons as the flow speed is increased, thus
rendering the basin of attraction of dendritic structures larger.

For small anisotropy and Prandtl number, we find oscillations of the
tip velocity in the presence of flow. On increasing the fluid
viscosity or decreasing the flow velocity, we observe a reduction in
the amplitude of these oscillations.

\end{abstract}



\bigskip
%{\bf Corresponding author:} \parbox[t]{9cm}{Klaus Kassner\\
%Otto-von-Guericke-Universität Magdeburg\\
%Postfach 4120 D-39016 Magdeburg\\
%Fax: +49-0391-6711205\\
\email{Klaus.Kassner@physik.uni-magdeburg.de}

%\vspace{1.5cm}
\maketitle

\section{Introduction}

Crystal growth from the melt or from solution almost never occurs in
convection-free conditions.  Notwithstanding this fact, models of
solidification often focus, when dealing with aspects of morphological
stability and pattern formation, on situations where convection is
absent \cite{muellerkrumbhaar91,billia93,%
  coriell93,glicksman93}.  Rather than by the negligibility of
convective effects, such a choice is generally motivated by the
difficulty of including them and the argument that the basic
prototypes of patterns appear and may be studied without convection.
This in turn has led to the tailoring of experiments on dendritic
growth, in which it was explicitly tried to avoid disturbing flows
\cite{glicksman93,glicksman94,lacombe95,bisang96}.

When convection was taken into account in calculations, it was usually
in simplified geometries or in conditions enabling simplification
of the model \cite{davis93,marietti2001a,marietti2001b}; often the
moving-boundary aspect of the problem was neglected
\cite{mueller94,lemarec97}.  Attempts to model the full problem
including convection and the motion of the liquid-solid interface were
essentially made only in cases where the deflection of the interface
remained relatively small \cite{kopetsch90,dupret94}.  The state of
the art a decade ago may be summarized roughly by saying that pattern
formation in crystal growth could be well simulated either for the
solid, treating the free-boundary problem in its full complexity, or
for the liquid, obtaining the convection roll pattern with good
accuracy by use of simplifying assumptions for interface motion.

With the advent of efficient phase-field techniques \cite{karma96b},
the solution of the moving-boundary problem became simpler, and first
simulations of convection in dendritic growth were performed in
diverse geometries and with both imposed flows and natural convection
\cite{toenhardt98,becker99,tong2001}.  In these, the Navier-Stokes
equations were solved by standard numerical approaches implementing
the partial differential equations either within a finite-element or a
finite-difference scheme. It was then a natural idea to supplement the
efficient approach to interface motion by an efficient non-iterative
method for flow simulation, the lattice-Boltzmann scheme.  This
approach was pioneered by Miller {\em et al.\/}
\cite{miller2001,miller2002,miller2003}, and a slightly different
variant, the advantages of which will be discussed in
Sec.~\ref{sec:model}, was developed by ourselves
\cite{dmitryjcg05,MKas2005b}.

Numerical studies of pattern formation solving the combined
free-boundary and flow problems will be useful in guiding the
development of analytic selection theory for dendritic growth and
other growth modes in the presence of convection.  At present, there
is a well-developed theory for purely diffusion-limited dendritic
growth, both in two and three dimensions
\cite{caroli86,benamar86,meiron86,kessler86,barbieri87,tanveer89,%
benamar93,brener93b}.
It provides an analytic demonstration of the existence of a discrete
set of needle crystal solutions to the model equations and shows that
the fastest of these solutions is linearly stable.

Acceptance of this microscopic solvability theory has not been
uncontroversial, as there is no clear agreement of its predictions
referring to crystalline anisotropy with experimental results
\cite{glicksman89}.  On the other hand, the mathematical statements of
the theory can hardly be disputed.  Therefore, the existence of a
needle crystal solution is a fact and its stability shows that it is
an attractor of the dynamics.  In principle, its basin of attraction
might be so small as to render it irrelevant experimentally.  But this
is essentially excluded by numerical simulations that have shown both
in two \cite{saito88,ihle94} and three \cite{karma98} dimensions that the
dynamical state of the system more or less automatically approaches
the prediction of selection theory.  In two dimensions at least, this
is also true for the dendrite decorated with side branches.  Numerics
and theory agree with each other so that experimental discrepancies
are most likely due to the fact that anisotropies in real crystals do
not correspond to the model expressions used in the theory or else
that additional effects interfere which are absent or controlled in
the simulations (for example, thermal fluctuations not considered in
the deterministic theory might affect the operating point of the
dendrite at low undercoolings \cite{pelce06}).

One of the more surprising predictions of microscopic solvability
theory was the nonexistence of dendrites in the absence of any kind of
anisotropy, be it that of surface tension or of interface kinetics.
Due to the nature of the theoretical approach, which is singular
perturbation theory about an Ivantsov parabola or paraboloid
\cite{ivantsov47}, such a statement can hold only for needle crystals
with a shape close to the (exact) solution of which they are supposed
to be small perturbations.  And it turned out later that indeed
steady-state crystal growth at zero anisotropy is possible, but only
with a shape that is far from an Ivantsov solution.  These new
structures were called doublons \cite{ihle94} in two dimensions and
selection theory has been developed for them as well \cite{benamar95}.
Since they continue to exist at finite anisotropy, there is a
coexistence range with dendrites, which means that there are two
attractors of the dynamics.  The standard argument is then that the
faster of the two morphologies will win, which in the analytically
tractable case is the doublon, whenever it exists.

Large scale two-dimensional structures consist of arrays of dendrites
or of doublons evolving in a noisy environment via side branching
and/or tip splitting processes.  A theoretical description of the
resulting dendrite and seaweed morphologies, based on scaling
arguments and selection theory \cite{brener96} gives a kinetic phase
diagram in the parameter space of undercooling versus surface tension
anisotropy.  All of the analytic predictions mentioned so far
refer to diffusion-limited growth only.

A first attempt to extend selection theory to situations with a flow
was made by Bouissou and Pelcé \cite{bouiss89b} and there were a
number of less rigorous approaches to the problem as well (references
are given in \cite{MKas2005b}).  However, while one experiment seems
to support this theory \cite{bouissou89}, another one contradicts it
\cite{lee93}.  Moreover, the theory is based on a linearization of the
basic equations, an approach that has been found not to always be
reliable \cite{tanveer2000}.  Clearly, more numerical or experimental
data are needed to both check the existing theories and guide further
theoretical development.  The purpose of this article is to provide
first elements of these data using our combined phase-field
lattice-Boltzmann approach \cite{dmitryjcg05,MKas2005b}.  A successful
selection theory on the microscale (constructed on the basis of these
and similar data) would then yield useful information for more applied
work on macrosegregation and related questions.

The paper is organized as follows. In Sec.~2, we discuss the basic
model equations and describe the method of their numerical
implementation.  In Sec.~3, we consider the influence a parallel flow
on the selection of growth velocity and tip radius, whereas in Sec.~4,
changes in the position of the morphology transition from dendrites to
doublons induced by the flow are discussed.  Some concluding remarks
are given in Sec.~5.

\section{Model equations and implementation \label{sec:model}}

For simplicity, we consider a symmetric model with equal thermal
diffusivities of the solid and liquid phases, expecting it to display
all the qualitative features of the more general case.  Moreover,
since we wish to confront our simulations with theoretical results on
different morphologies, we restrict ourselves to two-dimensional
systems here.  So far, there are very few results on non-dendritic
structures in three dimensions.

The well-accepted sharp-interface description of non-facetted crystal
growth from a supercooled melt in the presence of a fluid flow with
velocity ${\bf U}$ then starts from the following set of bulk and
interface equations:
\begin{eqnarray}
 \partial_t u +{\bf U}\nabla u &=& D\nabla^2 u, \nonumber\\
 {\bf n}\cdot {\bf V}&=& D{\bf n}\cdot(\nabla u|_s-\nabla u|_l),
 \label{eq:sh-int}\\
 u_i &=& -d(\theta)/R_K-\beta\, {\bf n}\cdot{\bf V}. \nonumber
\end{eqnarray}
Herein, $\,u=c_p(T-T_m)/L\,$ with $c_p$ denoting the heat capacity and
$L$ the latent heat, both per unit volume, is a nondimensionalized
temperature, $T$ being the temperature at the considered position and
$T_m$ the bulk melting temperature.  $D\,$ is the thermal diffusivity,
${\bf n}\,$ the local normal to the liquid-solid interface (pointing
from the solid into the liquid) and {\bf V} the interface velocity.
The subscripts of the temperature gradients in the second equation
refer to the solid and the liquid sides of the interface,
respectively.  $d(\theta)$ is the capillary length, related to the
orientation-dependent surface tension $\gamma(\theta)$ by
$d(\theta)=(d_0/\gamma_0)\,[\gamma(\theta)+\gamma''(\theta)]$.
$\theta$ is the angle between the interface normal and some fixed
direction (usually identified with the $x$ axis of the coordinate
system), $\gamma_0$ the average of the interface tension over all
orientations, and $d_0=\gamma_0 T_m c_p/L^2\,$ the similarly averaged
capillary length. $R_K\,$ is the local radius of curvature, $\beta\,$
the kinetic coefficient. In principle, $\beta\,$ is an orientation
dependent quantity just as the surface tension;  however, we will
restrict ourselves to the case of fast attachment kinetics here,
meaning that $\beta\,$ becomes negligible.  This implies certain
constraints on the choice of parameters of the phase-field model (see
below).
%\cite{MKas2005b}.

The boundary condition for the normal velocity is usually referred to as
Stefan condition, while the last equation in (\ref{eq:sh-int}),
describing the interface temperature, is the Gibbs-Thomson relation
(for $\beta=0$) with kinetic correction (for $\beta\ne0$).

At infinity, the temperature in the solid approaches $T_m$,
corresponding to $u=0$, whereas in the liquid, it takes on some value
$T_\infty < T_m$, corresponding to $u = - \Delta$. The quantity
$\Delta$ is denoted as the nondimensional undercooling.

In order not to complicate things unnecessarily, we assume the melt to
be an incompressible Newtonian fluid, governed by the appropriate
version of the Navier--Stokes equations, supplemented by boundary
conditions at the interface
\begin{eqnarray}
 \partial_t {\bf U}+{\bf U}\nabla{\bf U} &=&
        -\frac{\nabla P}{\rho}+\nu\,\nabla^2{\bf U}\>,
  \nonumber\\
 \nabla\cdot{\bf U}&=&0\>,  \label{eq:fluidmotion}\\
  {\bf U}_i&=&0\>, \nonumber
\end{eqnarray}
where equal mass densities $\rho$ have been assumed in the two phases,
$\nu$ is the kinematic viscosity, and $P$ denotes the pressure of the
liquid. ${\bf U}_i$ is the flow velocity at the interface.  The
boundary conditions correspond to no-slip conditions for the
tangential velocity and describe the fact that for equal densities of
the two phases, the liquid is neither sucked towards nor rejected from
the interface; hence, its normal velocity at the interface is zero in
the laboratory frame (the rest frame of the solid).

It is convenient to use dimensionless variables in the analysis of the
growth process.  The tip radius can be nondimensionalized using the
capillary length as $\,\bar{R}=R/d_0$, whereas flow and growth
velocities become nondimensional via normalization with the
characteristic velocity given by the ratio of the thermal diffusion
coefficient and the average capillary length, that is
$\,\bar{U}=Ud_0/D\,$ and $\,\bar{V}=Vd_0/D$.

The material properties are characterized by the anisotropy of the
surface energy or rather that of the capillary length and by the
Prandtl number $\,{\rm Pr}=\nu/D$; the flow is characterized by the
Reynolds number $\,{\rm Re}=UR/\nu$.

We employ a combined phase-field/lattice-Boltzmann scheme where
solidification is simulated with the phase-field model of Karma and
Rappel \cite{karma96,karma98}, and the flow of the liquid as well as
convective and diffusive heat transport are modeled using a
lattice-Boltzmann (LB) method. This means that the actual equations
simulated are not those given above but a phase-field approximation to
the interface dynamics (involving a finite-width transition region
between the solid and its melt) and a set of kinetic equations that
are asymptotically equivalent to the Navier-Stokes and
advection-diffusion equations.  

In particular, the phase-field equations read
\begin{eqnarray}
  \tau(\theta)\partial_t\psi &=&\left(\psi-
\lambda u\left(1-\psi^2\right)\right) \left(1-\psi^2\right)\nonumber\\
&& \mbox{} +\nabla\cdot\left(W^2(\theta)\nabla\psi\right)
 \nonumber \\
 && \mbox{}\quad -\partial_x\left(W(\theta)W'(\theta)\partial_y\psi\right) 
 \nonumber\\
 && \mbox{}\quad +
   \partial_y\left(W(\theta)W'(\theta)\partial_x\psi\right) \>, \label{e-pf}\\
  \partial_t u+{\bf U}\nabla u&=&D\nabla^2 u+
   \frac{1}{2}\partial_t\, h(\psi).  \nonumber
\end{eqnarray}
The value $\,\psi=1\,$ of the phase-field variable is chosen to
represent the solid phase, whereas $\psi=-1$ corresponds to the liquid
phase.  $W(\theta)\,$ is an anisotropic interface width,
$\tau(\theta)\,$ a relaxation time, and
$\,\theta={\mbox{arctan}}(\partial_y\psi/\partial_x\psi)$ is the angle
between the normal on a level set of $\psi$ and the $x$ axis. For the
level set given by $\psi=0$, this angle is the same as the angle
$\theta$ in Eq.~(\ref{eq:sh-int}), otherwise it provides an extension of the
definition of the latter into the bulk.  $\,h(\psi)$ describes the
coupling of the diffusion field to the phase field via latent heat
production.  This function was chosen as $\,h(\psi)=\psi$, which
appears to be computationally most efficient \cite{karma98}.

Via a suitable asymptotic expansion, the equations of the
sharp-interface model (\ref{eq:sh-int}) can be derived \cite{karma98},
with the following expressions for the capillary length and kinetic
coefficient:
\begin{eqnarray*}
d(\theta)&=&\frac{I}{\lambda
  J}\left(W(\theta)+\partial^2_{\theta}W(\theta)\right),\;\\
\beta(\theta)&=&\frac{I}{\lambda J}\frac{\tau(\theta)}{W(\theta)}\left(
  1-\lambda\frac{K+JF}{2I}\frac{W^2(\theta)}{\tau(\theta)}\right).
\end{eqnarray*}
These equations, first given by Karma and Rappel \cite{karma96}, have
been shown to be equivalent to a second-order accurate standard asymptotic
approximation \cite{karma98,almgren99}.

Requiring
$$
  W=W_0 A(\theta), \quad \tau=\tau_0 A^2(\theta), \quad
  \lambda=\frac{2ID\tau_0}{(K+JF)W_0^2}.
  $$
  we can impose a vanishing kinetic coefficient \cite{karma96}.
  For $\,h(\psi)=\psi$, the values of the phase-field specific
  coefficients are
$\,I=2\sqrt{2}/3$, $J=16/15$, %K=0.13604$,%
%\footnote{The exact value is
 $K=\sqrt{2}\left(\frac{188}{225}-\frac{16}{15}\ln 2\right)$,
%}
and $\,F=\sqrt{2}\ln 2$
\cite{karma96,karma98,MKas2005b}.  We use the anisotropy function
$$
  A(\theta)\equiv \frac{\gamma(\theta)}{\gamma_0}=
1+\frac{\alpha}{15}\cos 4\theta\>,
$$
leading to
\begin{equation}
 d=d_0(1-\alpha\cos4\theta)\>,  \label{eq:defd0}
\end{equation}
which is the usual model expression exhibiting four-fold symmetry.

Moreover, we set $\tau_0=1,\, W_0=1$.

The equation for the phase-field $\,\psi\,$ was discretized on a
uniform spatial lattice with grid spacing $\Delta x=0.4$, and it was
solved using the explicit Euler method with constant time step $\Delta t$
in the range 0.008--0.016.

Both the flow of the liquid and the heat transport are simulated using
the LBGK method (see \cite{ChenDoolen98}). Its main variables are
one-particle distribution functions $\,f_k$ defined on the nodes of a
regular spatial lattice. Different labels $\,k\,$ correspond to
different velocities $\,{\bf c}_k$ from a fixed finite set. In the
two-dimensional model used here, we employ nine velocities, namely,
$\,{\bf c}_0=(0,0)$, ${\bf
  c}_k=(\cos((k-1)\pi/2),\sin((k-1)\pi/2))\delta x/\delta t\,$ for
$\,k=1\ldots4$, and ${\bf c}_k=\sqrt{2}(\cos((k-1/2)\pi/2),
\sin((k-1/2)\pi/2))\delta x/\delta t\,$ for $\,k=5\ldots8$. Here,
$\delta x\,$ is the grid spacing, equal for both directions, $\delta
t\,$ is the time step.  The effect of making the velocities
proportional to $\delta x/\delta t$ is that nonzero velocities lead to
nearest neighbour and next-nearest neighbour sites of the square
lattice in one time-step, {\em i.e.\/}, only lattice point positions appear in
the dynamics, no interpolations are necessary.

Inside the LBE part of our simulations, the grid spacing and time step
are both formally rescaled to one, which is the reason why we have used a
different notation for them here from that in the phase-field part of
the simulation ($\delta x$ and $\delta t$ {\em vs.\/} $\Delta x$ and
$\Delta t$), although they are actually the same ``material'' quantities.

The evolution equation for $\,f_k\,$ is
\begin{equation}
  f_k(t+\delta t,{\bf x}+{\bf c}_k\delta t)=
    f_k(t,{\bf x})+\frac{f_k^{eq}-f_k}{\tau_f}\delta t.\label{e:LBGK}
\end{equation}
Distribution functions are advected (first term on the r.h.s.) and
undergo a relaxation to equilibrium values $\,f_k^{eq}\,$ which are,
as usual, taken to be expansions of Maxwellians up to second order
in the fluid velocity $\,{\bf U}\,$
\begin{equation}
  f_k^{eq}=\rho w_k\left(1+\frac{{\bf c}_k\cdot{\bf U}}{c_s^2}+\frac{({\bf c}_k
    \cdot{\bf U})^2}{2c_s^4}-\frac{U^2}{2c_s^2}\right),\label{e:feq}
\end{equation}
with $c_s$ having the physical meaning of an isothermal sound velocity.
The local fluid density is given by $\,\rho=\sum\limits_k
f_k=\sum\limits_k f_k^{eq}$, the flow velocity is $\,{\bf
  U}=\sum\limits_k f_k{\bf c}_k/\rho$, and the weight coefficients are
$\,w_0=4/9, w_{1-4}=1/9, w_{5-8}=1/36$. This form of the equilibrium
distribution functions ensures mass and momentum conservation and
provides the correct form of the momentum flux tensor
\cite{ChenDoolen98,QianDHumLall92}.

Performing a Chapman-Enskog expansion, one can derive from
(\ref{e:LBGK}) the continuity and Navier-Stokes equations
\cite{ChenDoolen98}, with kinematic viscosity
$\,\nu=c_s^2(\tau_f-\delta t/2)$. The isothermal sound velocity is
$\,c_s=\delta x/\sqrt{3}\delta t$, for small flow velocities the fluid
is almost incompressible (effects of compressibility are proportional
to $\,U^2/c_s^2$).

The influence of the growing pattern on the fluid flow was simulated
as proposed in \cite{becker99,tong2001}. An additional dissipative
force was introduced in partially filled regions
$$
 {\bf F}_d=-\rho\nu\frac{2g \phi^2}{W_0^2}{\bf U},
$$
where $g$=2.757 and $\phi=(1+\psi)/2$ is the solid fraction. This
provides the correct velocity boundary conditions at the diffuse
interface (see \cite{becker99,tong2001}), {\em i.e.\/}, the sharp-interface
limit of the velocity boundary conditions of
Eq.~(\ref{eq:fluidmotion}) is correctly reproduced.  The value of the
constant $\,g\,$ was obtained in \cite{becker99,tong2001} by an
asymptotic analysis of plane flow past the diffusive interface.

The action of forces on a liquid was simulated by the exact difference
method of \cite{kuper04b}. The term $\,\Delta f_k=f_k^{eq}(\rho,{\bf
  U}+ \Delta{\bf U})-f_k^{eq}(\rho,{\bf U})\,$ is added to the r.h.s.
of eq.~(\ref{e:LBGK}), where $\,\Delta{\bf U}={\bf F}\delta t/\rho\,$
is the velocity change due to action of force $\,{\bf F}\,$ during time
step $\,\delta t$.  This form of the change of distribution functions
is exact in the case where the distribution is equilibrium before the
action of the force (then after the action the distribution remains
equilibrium), whence the name of the method.  In the case of a
non-equilibrium initial state, this method is accurate to second order
in $\,\Delta{\bf U}$.  It is simple enough and valid for arbitrary
lattices and any dimension of space.  

The heat transport equation in (\ref{e-pf}) was treated in a very
similar way via the introduction of nine additional distribution
functions $N_k(t,{\bf x})$.  An extensive discussion of these
algorithmic details is presented in \cite{MKas2005b}.

\begin{figure}[h!]
\noindent \center
\noindent
\includegraphics[width=5.0cm]{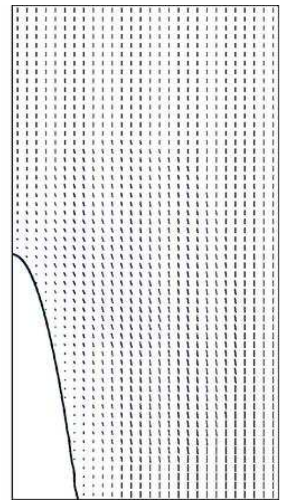}
\caption{Symmetric needle crystal, {\em i.e.\/},
  dendritic pattern in antiparallel flow.  Growth parameters:
  $\Delta=0.75$, $\alpha=0.45$, $\bar{U}=0.01$. The capillary length
  is $d_0=0.185$, the measured growth velocity $\bar{V}=0.0451$,
  leading to a diffusion length of 8.2 (20.5 lattice units).  The flow
  pattern is indicated by the streaks outside of the crystal.
  Numerical grid size: 700 $\times$ 1400 corresponding to
  1513.5$\,d_0$ $\times$ 3027.0$\,d_0$.
\label{f-dend}}

\end{figure}

\begin{figure}[h!]
\noindent \center
\noindent
\includegraphics[width=5.0cm]{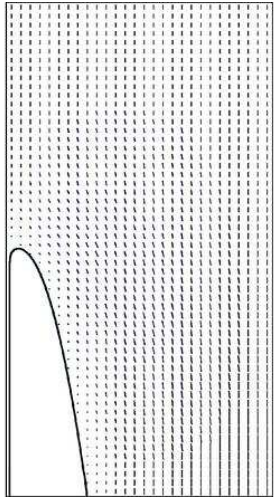}
\caption{Asymmetric needle crystal, (half of a) doublon
  pattern. Same growth parameters as in Fig.~\ref{f-dend}, except that
  $\alpha=0.3$. The capillary length and system size are the same as
  in Fig.~\ref{f-dend} as well, the measured growth velocity is
  $\bar{V}=0.0402$, leading to a diffusion length of 9.2, or 23
  lattice units.\label{f-doub}}
\end{figure}

In order to give a general impression of the type of results 
obtainable with these simulations, we display Figs.~\ref{f-dend} and
\ref{f-doub}, showing the steady states of a dendritic and a doublon
pattern, respectively.

A few remarks may be in order comparing our model with its main
predecessor as given by Miller {\em et al.\/}
\cite{miller2001,miller2002,miller2003}.  There are obvious
similarities, but our approach is simpler in the lattice-Boltzmann
part and has better convergence in the phase-field part.  The model
discussed in \cite{miller2001,miller2002,miller2003} is
four-dimensional and uses 24 velocities.  We have a two-dimensional
model with 2 $\times$ 9 velocities and our collision matrix is simpler.
So the lattice-Boltzmann part of our model is faster in two
dimensions, which is the only case considered here.

Concerning the phase-field part, our approach includes the Karma model
which has been shown to be quantitative at much larger interface
thicknesses than preceding alternatives. The phase-field model used by
Miller {\em et al.\/} has not been demonstrated to have any of the
advantages of the thin-interface asymptotics.  Its quantitative
accuracy might be challenged on the basis of the same objections as
that of the original Kobayashi model \cite{kobayashi94}.

Therefore, with the same accuracy prescribed, we expect the
phase-field part of our model to be much more efficient (because it
converges to the correct sharp-interface limit at much smaller system
sizes) than that of the model used in
\cite{miller2001,miller2002,miller2003} and the lattice-Boltzmann part
to be slightly more efficient.

\section{Selection of growth parameters}

The growth of a single needle crystal in parallel flow was simulated
for fixed surface tension anisotropy and a range of undercoolings and
flow velocities. To investigate the effects of parallel flow on the
growth characteristics, needle crystals grown without flow were used
as initial configurations. Reaching the steady state in the absence of
flow took between 90000 and 300000 time steps.

After loading the values of the temperature and phase fields, the flow
was initialized with boundary conditions of constant flow velocity
perpendicular to the upper boundary and zero velocity gradients at the
lower boundary, while the side boundaries were made reflecting. First, the
flow was allowed to evolve with a fixed configuration of the solid,
and the relative velocity error was calculated at each time step as
$$
  U_{err}=\frac{\sum |\hat{U}_x-U_x|+|\hat{U}_y-U_y|}{\sum |U_x|+|U_y|}.
$$
Here, $\hat{\bf U}\,$ refers to the flow velocity at the current, ${\bf
U}$ to that at the preceding time step, the summation is over all grid
nodes. The convergence condition was $\,U_{err}\le 10^{-5}$. 

Then the growth of the pattern was ``switched on'' and continued until
a steady, {\em i.e.\/}, constant-velocity, state was reached. In the range of
undercoolings $0.72 < \Delta < 0.76$, this took on the order of 150000
time steps.

%%% check
The numerical grid in these runs had a size of 700 $\times$ 1400,
corresponding to between 505$\,d_0$ $\times$ 1010$\,d_0$ and
2014$\,d_0$ $\times$ 4028$\,d_0$.  For the smallest measured
velocities, the diffusion length remained below 250 lattice units, for
the largest one, it was about 15 lattice units.  Therefore, in all
cases the system size was large enough to consider finite-size effects
negligible, in particular in view of the fact that the computational
domain corresponded to half the system size only (see
Figs.~\ref{f-dend} and \ref{f-doub}).

All the simulations discussed in this section were done either until
convergence of the pattern to a steady state was achieved or such a
steady state was found to be unattainable -- below we report on the
appearance of oscillatory states in certain parameter regions.  Only
then growth characteristics such as the growth velocity were measured,
{\em i.e.\/}, care was taken to avoid transient states providing only
information about temporary growth characteristics.

\begin{figure}[h!]
\noindent \center
\noindent
\includegraphics[width=7.7cm]{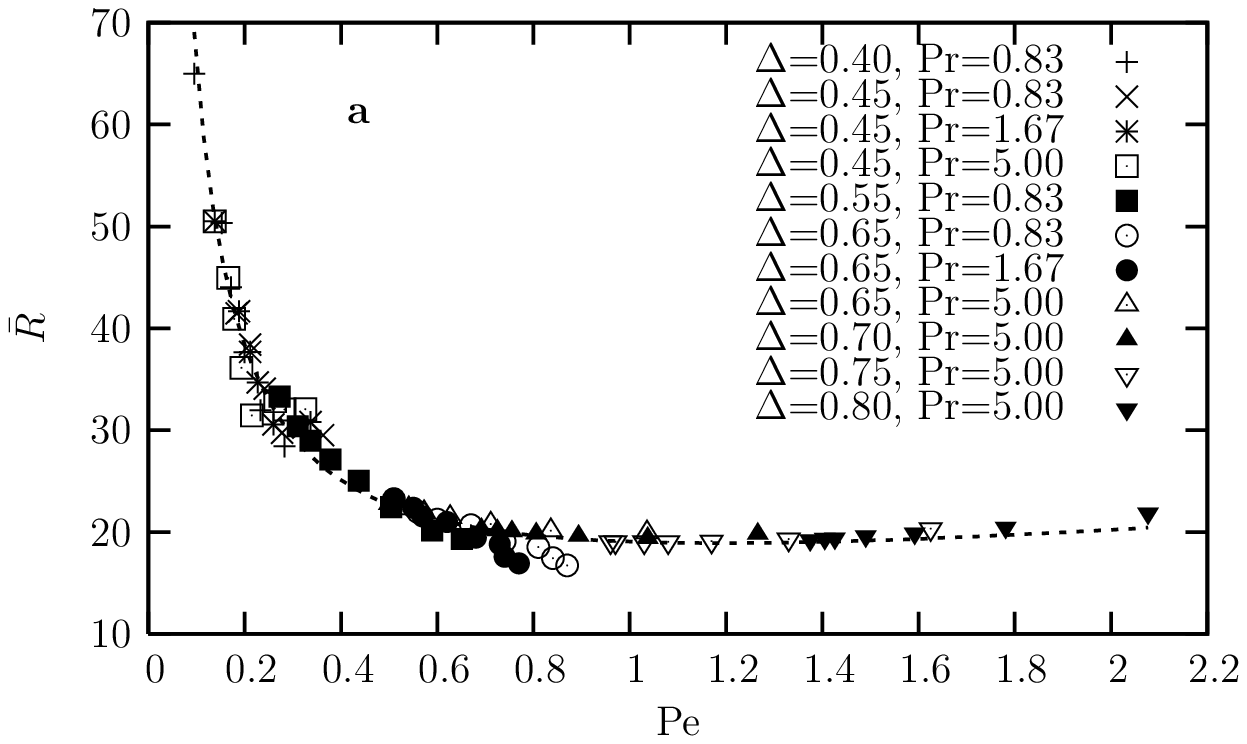}
\medskip\medskip\medskip
\noindent
\hspace*{-0.3cm}
\includegraphics[width=7.9cm]{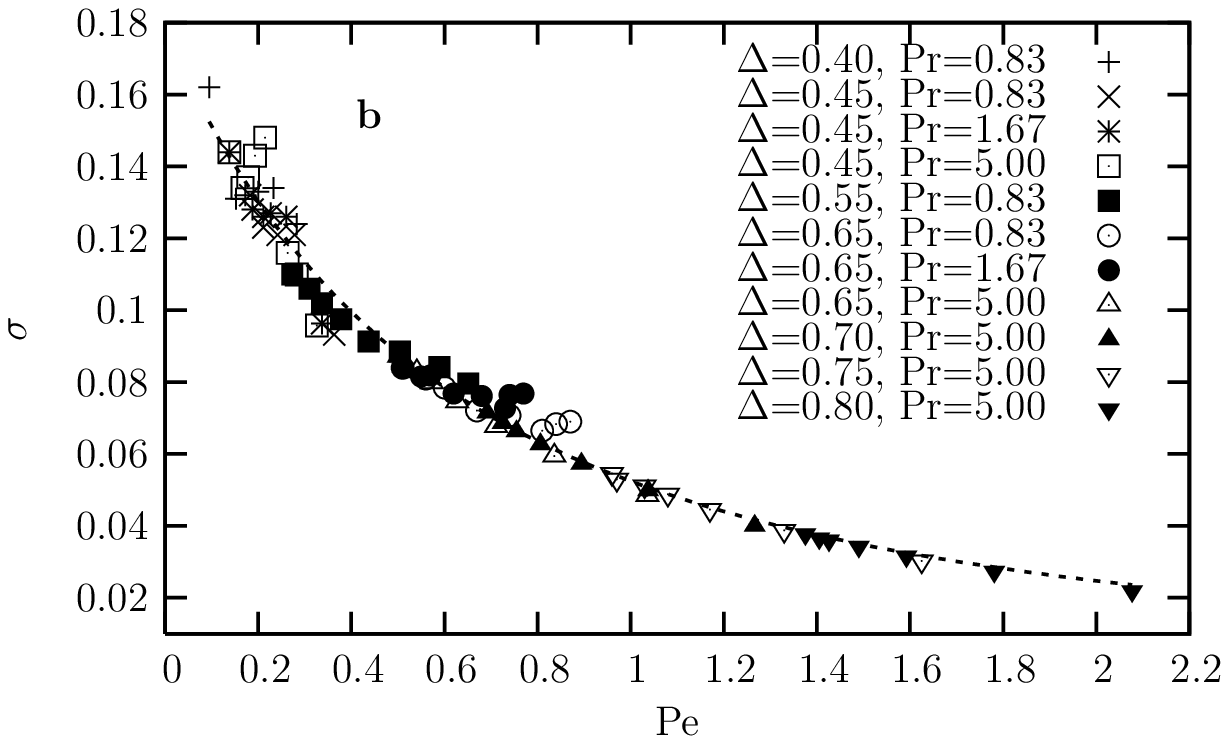}
\caption{
  Dependence of the tip radius (a) and selection parameter $\sigma$
  (b) on the P\'{e}clet number for dendrites.  Anisotropy parameter
  $\alpha=0.75$. Each symbol corresponds to the undercoolings and
  Prandtl numbers given in the legend and a value of the velocity
  $\bar{U}$ of the imposed flow in the range from 0 to 0.32. Larger
  values of $\bar{U}$ correspond to larger P\'{e}clet numbers.  The
  dashed line in (b) is a fit to $f({\rm Pe}) = a/\left(1+b\,{\rm
      Pe}\right)^2$, the dashed line in (a) is computed as $g({\rm
    Pe}) = 1/\left[f({\rm Pe}) \, {\rm Pe}\right]$.  From the fit, $a
  = 0.178$, $b = 0.841$.
\label{f-peclet}}
\end{figure}

Computed values of the reduced tip radius $\,\bar{R}\,$ and selection
parameter $\,\sigma=2Dd_0/R^2V=2/\bar{R}^2 \bar{V}\,$ are plotted
versus the growth P\'{e}clet number $\,{\rm
  Pe}=RV/2D=\bar{R}\bar{V}/2$ in Fig.~\ref{f-peclet}a and
Fig.~\ref{f-peclet}b for dendrites (single symmetric fingers).  In the
figure, the anisotropy of surface stiffness is $\alpha=0.75$, the
range of nondimensional initial undercoolings $\,\Delta=
c_p(T_m-T_\infty)/L\,$ extends from 0.4 to 0.8, and the reduced flow
velocity $\,\bar{U}=Ud_0/D\,$ is typically chosen between 0 and 0.32
(0, 0.01, 0.02, 0.04 etc.). One can see that for each of the two data
sets most of the points fall onto a unique curve. Minor deviations
appear mainly for small Prandtl numbers and large flow velocities.
The range of Reynolds numbers investigated in this data set was $0
\le {\rm Re} \le 5.6$.  It is possible to define a relative Reynolds
number based on the flow speed in the reference system attached to the
surface of the growing crystal.  This Reynolds number, which was
never zero, of course, extended up to $\approx$ 7.1.

To obtain the tip radius, we fitted a parabola to an extended region
about the tip.  This procedure does not yield an approximation to the
inverse curvature at the tip itself, because due to anisotropy, the
shape deviates from a parabola close to the tip \cite{tong2000}.
Rather it defines the tip radius by a global parabolic envelope
associated with energy conservation; the P\'{e}clet number ${\rm Pe}$
calculated from this tip radius corresponds to the P\'{e}clet number
used in selection theory and in the absence of flow reduces to the
Ivantsov value, defined by $\Delta = \sqrt{\pi {\rm Pe}} \,\exp({\rm
  Pe}) \,{\rm erfc}\left(\sqrt{{\rm Pe}}\right) $.

The tip radius initially decreases with the P\'{e}clet number but
later begins to increase again (for this anisotropy, the minimum is at
about $\,{\rm Pe}=1$), while the selection parameter $\,\sigma\,$
decreases in the whole range of P\'{e}clet numbers investigated. That
the tip radius increases for large undercooling can be easily
understood: in the limit $\,\Delta=1\,$, the solution should approach
a planar front with $\,\bar{R}=\infty$.  This argument is made more
quantitative below.

From the theoretical point of view, the most interesting feature of
these results is the existence of a master curve, on which the data
fall for a wide range of parameters.  For this feature (if it holds
generally) allows us to use the theory of dendritic growth without
convection to make predictions of selected velocities and tip radii in
the presence of a forced flow.  In the absence of flow, the growth
P\'{e}clet number depends only on the undercooling.  As soon as flow
is introduced, the P\'{e}clet number depends both on the undercooling
and the velocity of the imposed flow.  What Fig.~\ref{f-peclet}b then
tells us is essentially that no matter how we produce a given
P\'{e}clet number, we should expect the same selected value of
$\sigma$ at fixed anisotropy.  Hence, if we change both the flow
velocity and the undercooling in a way that keeps the P\'{e}clet
number constant, the growth speed and shape remain unchanged.  This
means that the case with flow can be mapped to the case without flow.
Of course, the problem of determining the P\'{e}clet number, for given
undercooling and flow velocity at infinity, is in itself a nontrivial
task.  In limiting cases (small external flow speed), it may be
approximated by the value obtained for an Ivantsov-type solution of an
Oseen approximation to the problem with flow.

\begin{figure}[h!]
\noindent \center
\noindent
\includegraphics[width=8.0cm]{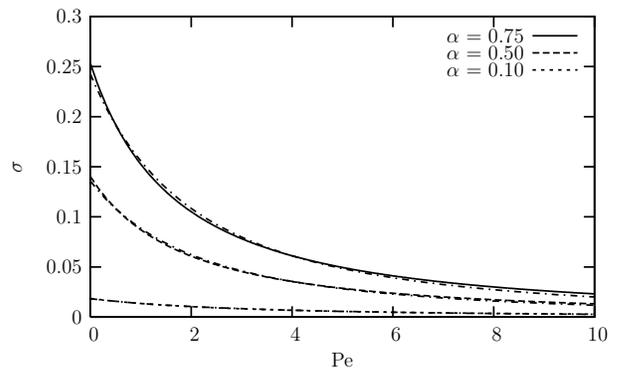}
\caption{ Behaviour of the selection parameter as a
  function of the P\'{e}clet number according to the linearized
  version of selection theory as given in \cite{barbieri89}.  The
  dash-dotted lines are fits with the same analytic expression as in
  Fig.~\ref{f-peclet}b.
\label{f-langer}}
\end{figure}

According to selection theory for the purely diffusion-limited case,
we should expect $\sigma$ to become independent of the P\'{e}clet
number for small anisotropy and small undercooling.  The latter
condition can be relaxed \cite{brener91} -- as long as ${\rm Pe}\>
\alpha^{1/2} \ll 1$, the standard result of selection theory, $V\sim
(D/d_0)\, \alpha^{7/4}\, {\rm Pe}^2$, continues to hold for a model
anisotropy of the type (\ref{eq:defd0}).  However, due to
computational limitations, this limit is difficult to access, hence
neither of these conditions is well satisfied in Fig.~\ref{f-peclet}b,
where $\alpha=0.75$ and ${\rm Pe}$ varies through 1.  The opposite
limit of large P\'{e}clet number is also known analytically
\cite{brener91}; the selection parameter should vary, for fixed
small anisotropy, proportional to $1/{\rm Pe}^2$.  Moreover, it is
possible to evaluate the predictions of solvability theory
\cite{barbieri89} numerically for arbitrary P\'{e}clet numbers.
Formally, this can be done for arbitrary values of the anisotropy
parameter $\alpha$ -- three examples are exhibited in
Fig.~\ref{f-langer} -- but the theory should not be expected to yield
good results for anisotropies that are not sufficiently small.

Comparing Fig.~\ref{f-peclet} with Fig.~\ref{f-langer}, we see that
the numerical results and the predictions from solvability theory
show the same general trend of $\sigma$ decreasing with increasing
P\'{e}clet number and that the order of magnitude of our $\sigma$
values is the same as for $\alpha = 0.75$ in solvability theory.
However, the selection parameter decreases much faster with increasing
P\'{e}clet number than in the theory (note the different scales of the
${\rm Pe}$ axes).  This may not be too surprising -- after all the
theory is made only for $\alpha \ll 1$, and while it has been claimed
to be quantitatively not too bad for $\alpha$ up to 0.5 or 0.6
\cite{barbieri89}, this claim referred to the small P\'{e}clet number
case.  Since the P\'{e}clet number appears only at next-to-leading
order in the small parameter of the theory ($\sqrt{\sigma}$), it is
perhaps not unreasonable to assume that the dependence of selected
characteristics on it is described less accurately within the theory
than, say, the anisotropy dependence at ${\rm Pe} = 0$.

Because the theory predicts the limiting behaviours of the selection
parameter at small and large P\'{e}clet numbers, it is tempting to try
to capture the behaviour at intermediate ${\rm Pe}$ by a simple
interpolating function.  The simplest rational function approaching a
constant value for ${\rm Pe}\to 0$ at finite slope and being
proportional to $1/{\rm Pe}^2$ at large ${\rm Pe}$ is $f({\rm Pe}) =
a/\left(1+b\,{\rm Pe}\right)^2$.  Fits with this function work pretty
well for both our numerical data and the results from selection
theory as is demonstrated in Figs.~\ref{f-peclet} and \ref{f-langer}.
Because $\bar{R} = 1/(\sigma {\rm Pe})$, this also yields a
description of the tip radii in Fig.~\ref{f-peclet}a as well as a
quantitative estimate for the tip radius behavior at large ${\rm Pe}$
($\sim {\rm Pe}\, b^2/a$).  The value of $a$ gives the limit of the
selection parameter for ${\rm Pe} = 0$, which in our case is about
30\% below the value obtained from selection theory (for an
illegitimately large value of $\alpha$).  

Note that given the graphs of $\sigma({\rm Pe})$ and $\bar{R}({\rm
  Pe})$, we could obtain a similar representation for the growth
velocity simply by plotting $\bar{V}=2/(\sigma \bar{R}^2)$ or, even
simpler, $\bar{V}=2\sigma {\rm Pe}^2$. Hence, the limiting growth
velocity for ${\rm Pe}\to\infty$ is $\bar{V} = 2 a/b^2$.

\begin{figure}[h!]
\noindent \center
\noindent
\includegraphics[width=7.7cm]{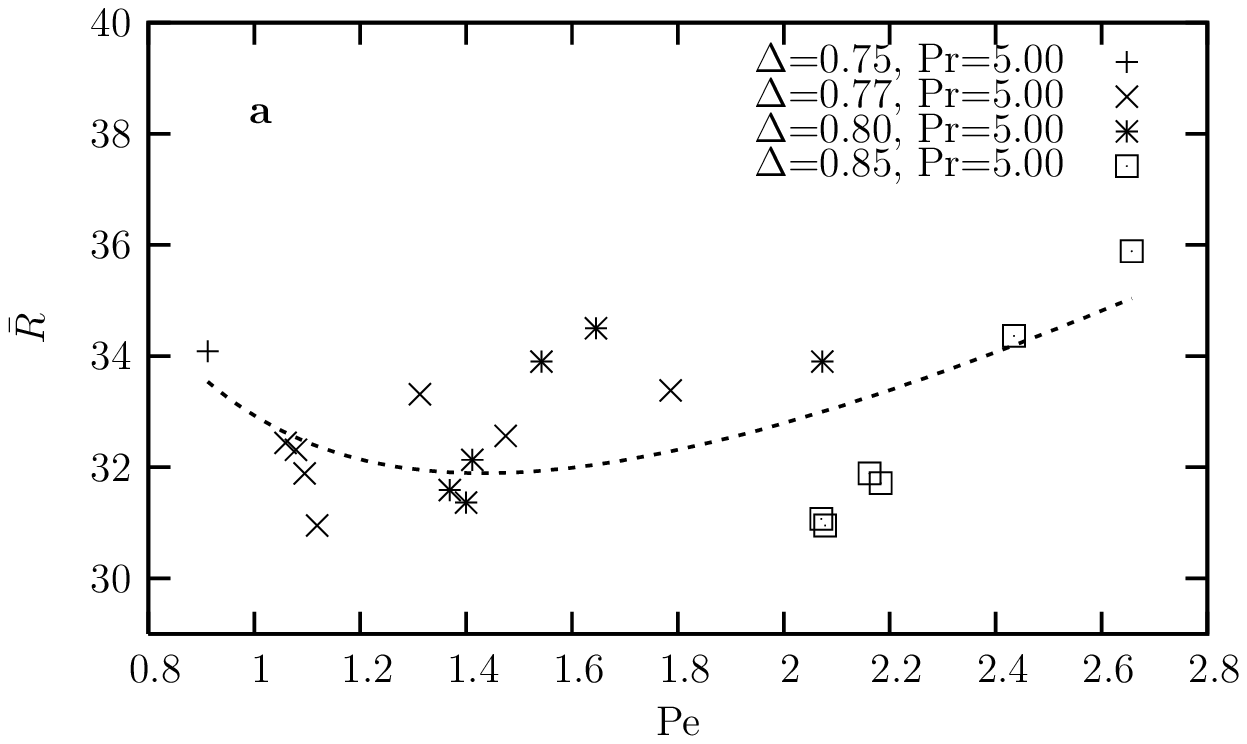}
\medskip\medskip\medskip
\noindent
\hspace*{-0.5cm}
\includegraphics[width=8.1cm]{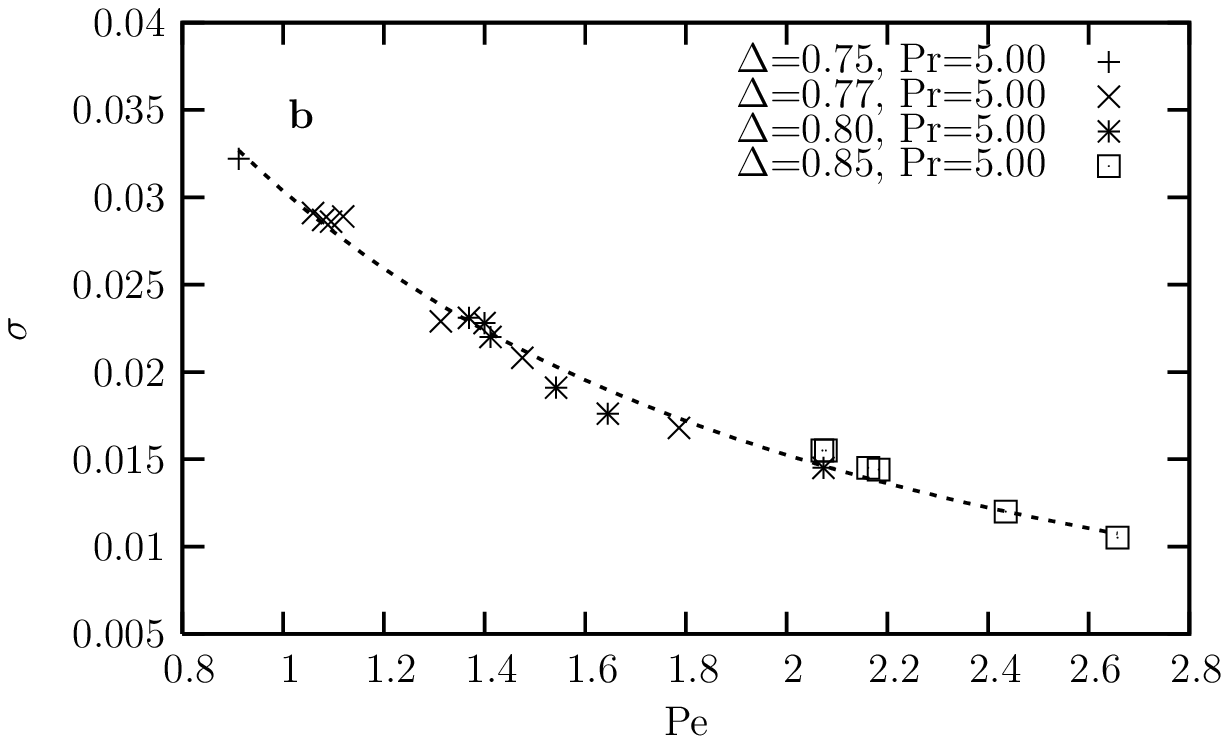}
\caption{Dependence of the tip radius (a) and selection parameter $\sigma$ (b)
  on the P\'{e}clet number for doublons.  Anisotropy parameter
  $\alpha=0.3$.  Appearance of the same symbol several times means
  different values for $\bar{U}$ (in the same range as in Fig.~1) at
  the same pair of values for the undercooling and the Prandtl number.
  The dashed lines are obtained by fitting as in Fig.~\ref{f-peclet},
  which yields $a = 0.0876$, $b = 0.699$.
\label{f-peclet2}}
\end{figure}

Results for doublons, {\em i.e.\/}, two asymmetric fingers with a
liquid-filled channel between them growing together, are presented in
Fig.~\ref{f-peclet2}. The surface stiffness anisotropy is 0.30 in this
case; the undercooling ranges from 0.77 to 0.85. For the reduced flow
velocity the same range from 0 to 0.32 was taken as for the symmetric
finger, whereas the Reynolds numbers extended only up to 2.1, as the
set of considered viscosities did not contain values as small as those
of Fig.~\ref{f-peclet}.

The ``tip'' radius was measured by fitting a parabola to the exterior
shape of the doublon, that is, only points much farther from the
central channel than the two tips of the pattern were used in the
fitting procedure. Since this procedure depends also on the cutoff
value defining which part of the shape is ``exterior'' and which one
is ``interior'', we do not expect a similar accuracy for this radius
as in the case of dendritic patterns.  Moreover, the total range of
radii displayed is about a factor of six smaller than in
Fig.~\ref{f-peclet}a, which contributes to making the results appear
much noisier than those for the dendritic pattern.

Nevertheless, while the characteristic length scale of the doublon may
not display the same clear-cut universality as that of the dendrite, a
unique dependence of the selection parameter on the P\'{e}clet number
is clearly visible in Fig.~\ref{f-peclet2}b.  A fit to the same
rational function $f({\rm Pe})$ as in the dendritic case is possible,
but less trustworthy than for the dendrites, as the range of
accessible P\'{e}clet numbers is smaller.  Moreover, its theoretical
foundation is weaker than for dendrites, as selection theory for
doublons does not yet seem to have been worked out in the limit of
large P\'{e}clet number.

\begin{figure}[h!]
\noindent \center
\includegraphics[width=8.5cm]{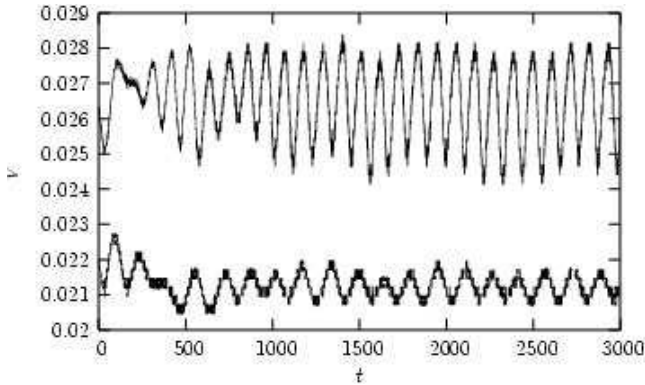}
\caption{ Measured growth velocity $\bar{V}$ of a
  dendrite as a function of time $t$.  $\Delta=0.7$, $\alpha=0.15$,
  $\bar{U}=0.04$, the upper curve corresponds to ${\rm Pr}=1.78$, the
  lower one to ${\rm Pr}=5.00$.  The rather strong flow provokes
  oscillations of the tip speed with large amplitude for small
  viscosity (${\rm Pr}=1.78$) and small amplitude for large viscosity
  (${\rm Pr}=5.00$).
\label{f-vel-osc}}
\end{figure}

Finally, it should be mentioned that the introduction of an external
flow may lead to a loss of stability of steady-state growth and result
in instationary patterns.  With small anisotropy and Prandtl number,
oscillations of the tip velocity are observed.  This observation may
relate to the prediction by the selection theory presented in
\cite{bouiss89b} that above a certain flow velocity no steady-state
solutions will be possible any more. Increase of the fluid viscosity
and/or decrease of flow velocity damps these oscillations as shown in
Fig.~\ref{f-vel-osc}.

From the existence of these oscillations, it may be concluded that
there are parameter regions (attained for given anisotropy on
decreasing the Prandtl number) where the simple picture discussed
above breaks down.  Selected growth characteristics will then not
depend on the growth P\'{e}clet number and the anisotropy parameter
alone.  For these oscillations are not predicted by solvability theory
without flow, hence the simple mapping to this theory is not feasible
anymore, and an extension of selection theory such as the one given in
\cite{bouiss89b}, but preferably on a more rigorous basis, becomes
necessary.

\section{Morphology diagram}

Previously, a kinetic phase diagram was obtained theoretically
\cite{brener96} in the case of purely diffusive growth, distinguishing
four morphologies: compact dendritic structures at large anisotropy
and not too large undercooling, compact seaweed patterns at large
undercooling (and not too large anisotropy), noise-dominated fractal
dendritic and seaweed morphologies at sufficiently small anisotropy
and undercooling, respectively. The transition lines between the
different morphologies and their nature (as first- or second-order
kinetic phase transition or cross-over) were determined analytically
under certain limit assumptions.  Regarding the compact growth
morphologies, it was stated that doublons cease to exist for larger
anisotropies, but when they exist, they are faster than dendrites.  In
principle, the latter exist at all nonzero anisotropies, but they are
overtaken and thus overgrown by doublons in the region of coexistence
(hence the transition from compact dendrite to compact seaweed would
be first order, because both morphologies coexist above a certain
undercooling).

How an imposed external flow may influence the different growth
patterns is interesting and largely unexplored.  We have already
shown that doublons survive in a shear flow
\cite{dmitryjcg05,MKas2005b}, a somewhat counterintuitive result.

In the present work, we investigate the morphology diagram for growth
in a parallel flow imposing a number of different flow velocities,
with a particular view to the positions of the transition lines
between doublon and dendrite growth.

\begin{figure}[h!]
\noindent \center
\includegraphics[width=9.0cm]{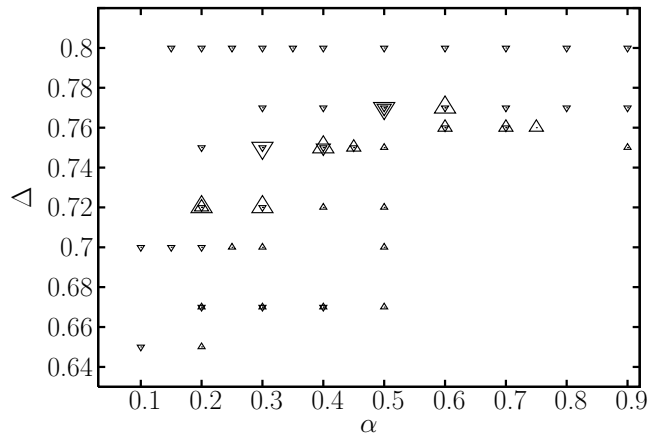}
\caption{ Morphology diagram displaying the
  predominance of dendrites or doublons at different flow speeds.
  Triangles correspond to dendrites either being the only morphology
  or the faster one, inverted triangles correspond to doublons being
  faster.  There are three sizes of symbols.  The smallest triangles
  refer to the case without flow described in current analytic
  theories.  Medium-size triangles are for a flow speed
  $\bar{U}=0.01$; big ones for a flow speed $\bar{U}=0.04$.  The
  general trend is that with increasing flow speed dendrites invade
  the original domain of doublons.  There are three points at
  $\Delta=0.67$, where the simulation gave the same velocities for
  both structures to three significant digits.
\label{f-morph-u000_u004}}
\end{figure}

Figure~\ref{f-morph-u000_u004} gives an overview of the measured
morphology diagram (actually a small section only of the entire plane
undercooling versus anisotropy) for the purely diffusive case and two
different flow velocities.  In relating this to previously measured
transitions between the dendritic and doublon morphologies
\cite{ihle94,kupferman95} (at zero flow), it should be kept in mind
that these older numerical results refer to the {\em one-sided
  model\/} whereas here we consider the {\em symmetric\/} model.  As
it turns out, the transition line is shifted to higher values of the
anisotropy ({\em e.g.\/}~$\alpha$ between 0.20 and 0.25 at
$\Delta=0.7$ instead of $\alpha\approx 0.12$), which seems plausible,
because the added diffusion in the solid tends to reduce
anisotropy-induced temperature differences imposed at the interface.
In fact, phase-field simulations of the symmetric model (in the absence
of flow) by Tokunaga and Sakaguchi \cite{tokunaga2004} also exhibit
this shift to higher anisotropies. Their calculations were done in a
channel that is narrow in comparison with our system width, so they
introduced an intermediate morphology between doublons and dendrites,
two competing Saffman-Taylor \cite{saffman58,combescot86} like fingers
(ST) [the existence range of which should vanish for infinite system
size].  For $\Delta=0.7$, they find the transition from doublons to ST
at $\alpha \approx 0.24$, that from ST to dendrites at $\alpha \approx
0.32$, which agrees well with our result.

Because we simulate only one of the two fingers of a doublon, imposing
mirror symmetry about the system boundary (see Fig.~\ref{f-doub}), our
calculation suppresses possible antisymmetric instabilities of a
doublon, {\em e.g.\/}, instabilities, where one finger gets ahead of
the other.  However, there is some evidence \cite{ihle94} that on
increase of the anisotropy parameter doublons normally get unstable by
dynamical unbinding of the two fingers, which move apart and become
independent dendrites.  This unbinding instability is symmetric and
would not be missed by our approach.  All our statements about {\em
  existence\/} of doublons are, of course, not affected by the
possibility of an unstable mode not taken into account.  And finally,
we base our assertions about the predominance of one of the two growth
modes on comparisons of the velocities of both, which will come out
correctly of the computation with the imposed symmetry.  The worst
that could happen is that a doublon found to be faster than a dendrite
at the same parameter values is unstable with respect to an
antisymmetric perturbation, in which case the dendritic morphology
would survive, if it is stable.  Such a scenario is not very likely,
given the fact that our doublons, whenever they were faster than the
associated dendrites, exhibited closely-spaced tips, corresponding to
the predictions of selection theory \cite{benamar95}.

The case of purely diffusive growth is depicted in
Fig.~\ref{f-morph-u000_u004} by the smallest symbols.  Triangles with
their tips pointing upward correspond to dendrites, inverted triangles
to doublons.  On increase of the reduced flow velocity $\,\bar{U}\,$
to 0.01, denoted by larger triangles, dendrites become faster than
doublons at several combinations of undercooling and anisotropy.  The
largest triangles in Fig.~\ref{f-morph-u000_u004} correspond to a
velocity of $\,\bar{U}=0.04$.  They demonstrate how the region where
dendrites are faster than doublons increases with increasing flow
velocity.  

It should be noted that according to the analytic theory
\cite{benamar95} for the purely diffusive case doublons would always
be faster than dendrites at coexistence.  Dendrites would dominate
only where doublons did not exist. This is not quite true at the large
anisotropy values considered here.  For example, at the point
$\alpha=0.5$, $\Delta=0.67$, where we have put a symbol denoting
dendritic growth, doublons exist, too, but are slower than dendrites.
Nevertheless, it is rather remarkable that external flow can lead to
dendrites becoming sufficiently fast to outrun doublons in an extended
range of parameters.

Note that the morphology diagram should actually be displayed in three
dimensions, as it is spanned by the three variables $\alpha$,
$\Delta$, and $\bar{U}$.  We circumvent the need for a genuine 3D
representation by taking different symbol sizes to represent different
flows, as only few flow velocity values could be studied.

That the presence of a parallel fluid flow in general favours
dendrites over doublons is an additional possible reason for the
difficulty to obtain doublons in experiments, the main reason of
course being that in experiments the value of the undercooling is
usually so small that one is far from the existence region of
doublons.  Experimental approaches to produce doublons in crystal
growth either had to use artifices to obtain effectively vanishing
anisotropy in the growth plane \cite{akamatsu95} or led to the
observation of transient doublons only \cite{stalder2001,singer2004d}
-- these were however true 3D structures.

\section{Conclusions}

In summary, we use a previously proposed combined
phase-field/lattice-Boltzmann scheme to simulate dendritic growth from
a supercooled melt in external counterflows directed parallel to the
growing needle crystal. Several regions of the morphology diagram in
the space spanned by the anisotropy parameter, the nondimensional
undercooling and the nondimensional flow velocity were explored.

For dendrites at moderate to high undercooling and high anisotropy, we
found that the values of tip radius and selection parameter, and hence
of the growth velocity, depend on the growth P\'{e}clet number only,
not on the undercooling and flow velocity separately.  Hence, it may
be argued that the essential effect of a parallel flow, at least in a
certain part of parameter space, is to change the selected tip
radius and growth velocity solely by modifying (increasing) the
P\'{e}clet number. In this region, selection theory for the purely
diffusive case is applicable, the main task being to determine the
relationship between undercooling, imposed flow velocity and the
growth P\'{e}clet number. For doublons a similar dependence for the
selection characteristics was obtained.

Incorporation of genuine flow effects into selection theory does
become necessary as the anisotropy and Prandtl number become small,
when tip oscillations take over and the steady state either ceases to
exist or becomes unstable. Increase of the fluid viscosity and/or
decrease of flow velocity is observed to damp down these oscillations.

For smaller anisotropy, an interesting phenomenon is observed. The
growth velocity for dendrites increases faster than for doublons
with increase of the flow velocity (at the same undercooling and
anisotropy). For some parameters, dendrites become faster, hence,
external flow can lead to morphology transitions and change the
kinetic phase diagram.

{\bf Acknowlegments} Financial support of this work by the German
Research Foundation (DFG) under grant no. FOR 301/2-1 within the
framework of the research group ``Interface dynamics in pattern forming
processes'' is gratefully acknowledged.

\bibliography{crystal,lattice}

%\begin{figure}[h!]
%\noindent \center
%\includegraphics[width=12cm]{channel.eps}

%\caption{\label{ch-scheme}}
%\end{figure}

%\newpage

%\begin{figure}[h!]
%\noindent \center
%\includegraphics[width=14cm]{u.eps}

%\caption{\label{ch-shape}}
%\end{figure}

\end{document}